\documentstyle[12pt,epsf]{article}
\begin{document}
\font \gothic=eufm10  scaled \magstep1
\font \gothind=eufm7
\newcommand\goth[1]{{\mbox{{\gothic #1}}}}
\newcommand\gots[1]{{\mbox{{\gothind #1}}}}
\makeatletter
\def\figurename{Fig.}
\long\def\@makecaption#1#2{%
   \vskip 10\p@
   \setbox\@tempboxa\hbox{#1. #2}%
   \ifdim \wd\@tempboxa >\hsize
     #1. #2\par
     \else
       \hbox to\hsize{\hfil\box\@tempboxa\hfil}%
   \fi}
\makeatother

\begin{center}
{\LARGE\it  Spontaneous Compactification \\
to Robertson-Walker  Universe \\
Due to Dynamical Torsion

}

\vspace{8mm}
{\large\bf Viktoria Malyshenko and Domingo  Mar\'\i n Ricoy}\\
{\small vika@domar.pvt.msu.su}\hskip1.5cm {\small marin@domar.pvt.msu.su}\\[.2cm]
Moscow State University

\end{center}

\vspace{4mm}

\abstract
We investigate multidimensional gravity with the Gauss--Bonnet term and with
torsion on the space of extra dimensions chosen to be the group manifold of
a simple Lie group. We take the Robertson-Walker ansatz for the 4-dimensional
space-time and study the potential of a dilaton and torsion fields. It is
shown that for  certain values of the parameters of the initial theory the
potential has classically stable minima, corresponding to the spontaneous
compactification of the extra dimensions. However, these minima have zero
torsion.

\vspace{4mm}

\section{Introduction}
It is a common belief that space-time may have more than four dimensions.
Theories based on this assumption are often referred to as Kaluza-Klein
theories, since the initial idea was put forward by T.\,Kaluza and O.\,Klein in
1921 \cite{kal}. For additional dimensions not to contradict the observable
reality they must be small enough so that they cannot be probed in modern
experiments.  Usually, it is assumed that the multidimensional space-time has
the structure $E=M^{(4)}\times S$, with $M^{(4)}$ being the four dimensional
space-time and $S$ being the internal space of extra dimensions with a
characteristic size of the order of the Plank scale $L_{pl}\approx 10^{-33}$
cm.  In so doing, it is supposed that the direct product structure of the
multidimensional space-time $E=M^{(4)}\times S$ appeared dynamically, as a
result of some symmetry breaking at a very early stage of the cosmological
evolution of the universe.

However, investigations within the framework of pure multidimensional
Einstein gravity have shown that this theory is plagued with many serious
difficulties, which leads to the fact that the effective 4-dimensional theory
is hard to interpret. For example, there is no acceptable vacuum solution of
the form $E=M^{(4)}\times K/H$, where $M^{(4)}$ is Minkowski space-time and
$K/H$ is a homogeneous space with non-abelian isometry group $K$ (which is
necessary to obtain non-abelian gauge fields on $M^{(4)}$ \cite{app}). To
solve this problem one can add Yang-Mills or matter fields to the initial
theory from the very beginning (see, e.g. \cite{sal}), but this would be a
rather strong deviation from the original Kaluza-Klein idea. To remain closer
to Kaluza-Klein ansatz one can consider some generalized gravitational
theories, for example, theories with torsion.

An attractive feature of Kaluza-Klein theories is that scalar fields, which
are necessary, for example, to ensure the inflation in four dimensions or
spontaneous symmetry breaking mechanism, appear quite naturally in four
dimensions within this framework. They emerge from extra components of the
metric and can also be induced by torsion on internal space.  The potential
of these fields is determined by the geometry and symmetries of the
multidimensional space-time.

Compactification to manifolds with torsion has been investigated in many
papers (see, e.g.\cite{rud} and references therein).  It is well known that
torsion is not a dynamical variable in pure Einstein-Cartan gravity
\cite{tra}.  In the presence of matter, its dynamics is carried by the spin
density of matter. This property is no longer preserved if one adds higher
order curvature terms to the standard Einstein-Cartan Lagrangian. There are
many reasons for choosing the Gauss-Bonnet term ${\cal R}^2 =
R_{ABCD}R^{CDAB} - 4R_{AB}R^{BA} + R^2$. For example, as it was shown in
\cite{mue}, \cite{lov}, if one adds such term to the initial Lagrangian, the
equations of motion include the derivatives of the metric  not higher than of
the second order therefore such teories are ghost-free.  Such form of the
curvature squared term is also motivated by quantum field theory limit of the
heterotic string models \cite{zwi}.  Besides, the presence of second order
curvature terms could explain the inflation of the observable Universe (see
e.g.  \cite{lin}).

In this paper we will investigate multidimensional gravity with torsion. The
ground state is chosen to have the structure of the direct product of
Robertson-Walker universe and a group manifold $S$ represented as a
homogeneous space $S=K/H$ with $K=S\times S$ being the isometry group and
$H=\mathop{\rm diag}\{S\times S\}$ being the isotropy subgroup.

The paper is organized as follows. In the next section we will give a brief
description of $K$~-~invariant metric connections with torsion on group
manifolds. In section 3 we calculate the components of the multidimensional
curvature tensor $ R^A_{BCD}$ and carry out the dimensional reduction of the
initial action. Section 4 is devoted to the effective scalar fields
potential. We will analyze its minima and consider possible cosmological
consequences of the model.

\section{Invariant connections with torsion on group manifolds}

In this section we briefly present some results from our previous paper
\cite{kub} with the purpose to make the discussion complete.

It is well known (see, e.g. \cite{kob}) that the group manifold of a simple
Lie group~$S$\break $(\mathop{\rm dim} S= d)$ can be represented as a reductive homogeneous
space $S=K/H$ with $K=S\times S$ being the isometry group and
$H=\mathop{\rm diag}\,(S\times S)$ being the isotropy subgroup.  We consider the class of
$K$-invariant metrics $g$ on $S$ (that is metrics which are invariant under
both right and left action of the group $S$) and metric connections with
torsion $\omega$ on the principal fiber bundle $O(S)$ of the orthonormal
frames over $S$. The group $K$ acts transitively on the base $S=K/H$ by left
multiplication of cosets and induces a natural automorphism of the bundle
$O(S)$.  The lift of the $K$-action to the bundle automorphism is
characterized by the homomorphism $\lambda: H \rightarrow SO(d)$.  We would
like to remind that if torsion is present then connection cannot be expresed
only in terms of derivatives of metric.  Let $\goth k$ be the Lie algebra of
the group~$K$, so that $\goth k = \cal S \oplus \cal S$ and ${\goth
h}=\{{(X,\,X)},\, X\in \cal S$\} (gothic letters stand for the corresponding
Lie algebras). The Lie algebra~$\goth k$ admits three natural reductive
decompositions:  ${\goth k} = {\goth h} \oplus {\goth m}$ with ${\goth m}=
{\goth m}_0,\; \goth m_+$ and $\goth m_-$
\begin{eqnarray} {\goth
m}_0\!\!&=\!\!& \Bigl\{( X/2, - {X}/2), \quad {X}\in {\cal S} \Bigr\}, \quad
(0)\; \mbox{connection}, \\ \nonumber {\goth m}_+\!\!&=\!\!& \Bigl\{(0, -
{X}), \quad {X}\in \cal S \Bigr\}, \quad (-) \;\mbox{connection}, \\
\nonumber {\goth m}_-\!\!&=\!\!& \Bigl\{({X}, 0), \quad {X}\in \cal S
\Bigr\}, \quad (+) \;\mbox{connection}.  \nonumber\end{eqnarray}

We denote by $o\in K/H$ the origin in $K/H$.  It is easy to see that there is
a natural isomorphism between the spaces $T_{o}(K/H),\; \goth m$ and $ R^d$,
that is $T_{o}(K/H)\cong \goth m\cong R^d$.  All $K$-invariant connections
$\omega$ on the bundle $O(S)$ are given by Wang's theorem \cite{kob}.  It
states that there is a 1-1 correspondence between the $K$-invariant
connections on the bundle $O(S)$ and linear mappings $\Lambda$
$$
\Lambda: {\goth m} \rightarrow {\cal SO}(d) \equiv Lie \bigl(SO(d)\bigr)
$$
such that
$$
\Lambda \Bigl(\mathop{\rm ad}  h(\tilde X)\Bigr) =
\mathop{\rm ad} \Bigl(\lambda(h)\Bigr)\Bigl(\Lambda(\tilde X)\Bigr),
\qquad \tilde X\in {\goth m},\quad h\in H.
\eqno(2)
$$
In terms of the mapping $\Lambda$ the formulas for the invariant torsion $T$
and curvature $R$ on $K/H$ at the point $o$ take the form
$$
T_o(\tilde X,\tilde Y) = \Lambda(\tilde  X)\tilde  Y
-\Lambda(\tilde  Y)\tilde  X -[\tilde  X,\tilde Y]_{\goth m},
\eqno(3)
$$
$$
R_o(\tilde X,\tilde Y) = [\Lambda(\tilde X),\Lambda(\tilde Y)] -
\Lambda([\tilde X,\tilde Y]_{\goth m}) - \lambda([\tilde X,\tilde Y]_{\goth
h}),  \qquad \tilde X,\tilde Y \in \goth m.  \eqno(4)
$$

Since $SO(d) \cong R^d \wedge {R^d}$ we can introduce the mapping $\beta :
{\goth m}\otimes {\goth m}\rightarrow \goth m$
$$
\beta(\tilde X,\tilde Y) \equiv \Lambda(\tilde X)\tilde Y.
$$
Then we decompose the connection form $\omega$ into a sum of the Levi-Civita
connection $\stackrel{\circ}{\omega}$ and the so-called contorsion form
$\bar\omega$.  Consequently, for the mappings $\Lambda$ and $\beta$ we have
$\Lambda = \stackrel{\circ}{\Lambda} +\bar {\Lambda} $ and
$\beta=\stackrel{\circ}{\beta}+\bar\beta$. The expression for
$\stackrel{\circ}{\Lambda}$ was obtained by Nomizu \cite{kob}
$$
\stackrel{\circ}{\beta}(\tilde X,\tilde Y) \equiv
\stackrel{\circ}{\Lambda}(\tilde X)\tilde Y = \frac{1}{2}[\tilde X,\tilde
Y]_{\goth m} + \stackrel{\circ}{U}(\tilde X,\tilde Y), \quad \tilde
X,\,\tilde Y \in \goth m,
$$
where $\stackrel{\circ}{U}(\tilde X,\tilde Y)$ is some symmetric bilinear
mapping ${\goth m}\stackrel{s}{\otimes} {\goth m}\rightarrow \goth m$. We
will discuss it later.

It is also useful to decompose $\bar\beta$, which describes contorsion, into
symmetric and antisymmetric parts $\bar\beta= \bar\beta_s + \bar\beta_{as}$.

Hence
$$
\beta_{as}=\frac{1}{2}[\tilde X,\tilde Y]_{\goth m}
+ \bar\beta_{as}(\tilde X,\tilde Y),
\eqno(5)
$$
$$
\beta_s=\stackrel{\circ}{U}(\tilde X,\tilde Y)
+ \bar\beta_s(\tilde X,\tilde Y).
\eqno(6)
$$

It is well known \cite{kob} that the $K$-invariant metrics $\gamma$ on $K/H$
are in 1-1 correspondence with non-degenerated symmetric  $\mathop{\rm ad} H$-invariant
bilinear forms $B$ on $\goth m$, i.e.  $B(\tilde X,\tilde Y) =
\gamma_{o}(\tilde X,\tilde Y),\;$  $ \tilde X,\,\tilde Y\!\in \!{\goth m }$.
The invariance of $B$ with respect to $\mathop{\rm ad}  H$ means that
$$
B\Bigl([\tilde A,\tilde X],\tilde Y\Bigr) +B\Bigl(\tilde X,[\tilde A,\tilde
Y]\Bigr) =0,\; \qquad \tilde X,\tilde Y \in {\goth m}, \quad \tilde A\in
{\goth h}.
\eqno(7)
$$
In can be shown that in terms of $\beta$ the metricity
condition reads
$$
B\,\Bigl(\beta\,(\tilde X,\tilde Y)\,,\,\tilde Z\Bigr)
+B\,\Bigl(\beta\,(\tilde X,\tilde Z)\,,\,\tilde Y\Bigr)=0,
\quad \tilde X,\;\tilde Y \in \goth m.
$$
Combining this condition  with two other formulas obtained from it by the
cyclic permutation of $\tilde X,\;\tilde Y$ and $\tilde Y$, it is easy to
derive the following relation between the symmetric part $\beta_s$ and the
antisymmetric part $\beta_{as}$ of the full mapping $\beta(\tilde X,\tilde
Y)$
$$
B\,\Bigl(\beta_s\,(\tilde X,\tilde Y)\,,\,\tilde Z\Bigr)
= B\,\Bigl(\beta_{as}\,(\tilde Y,\tilde Z)\,,\,\tilde X\Bigr) +
B\,\Bigl(\beta_{as}(\tilde Z,\tilde X)\,,\,\tilde Y\Bigr),
\quad \tilde X,\;\tilde Y \in \goth m,
\eqno(8)
$$
which  enables us to calculate the symmetric part through the
antisymmetric one. So, our next step will be to find the operator
$\bar\beta_{as}$.

Let us write down
the invariance condition (2) in the infinitesimal form
$$
\bar\beta_{as}\Bigl((\mathop{\rm Ad} \rho \wedge 1 + 1\wedge \mathop{\rm Ad} \rho)\,\xi\Bigr)
= \mathop{\rm Ad} \rho\,\Bigl(\bar\beta_{as}\,(\xi)\Bigr),
\qquad \xi \in \goth m\wedge {\goth m},\quad \rho\in \goth h.
$$
Now we see that $\bar\beta_{as}$ can be interpreted as an operator which
intertwines the equivalent representations of the algebra $\goth h$ in the
linear spaces $\goth m\wedge \goth m$ and $\goth m$. To construct this
operator explicitly we use the general method applied to the dimensional
reduction of the invariant connections in multidimensional Yang-Mills
theories (see e.g. \cite{vol}).

It is evident that $\goth m\cong \goth h \cong {\cal S}$ and the linear
spaces $\goth h$ and $\goth m$ carry the adjoint representation of the
algebra $\cal S$. So, it is necessary to decompose the antisymmetrized tensor
product $\mathop{\rm ad} {\cal S} \wedge \mathop{\rm ad} {\cal S}$ into irreducible representations
(irreps) of the algebra $H$ and to check, whether  there is the adjoint
representation $\mathop{\rm ad} \cal S$ among them. In \cite {kub} it has been
demonstrated that for the simple classical Lie algebras the decomposition of
$\mathop{\rm ad} {\cal S} \wedge \mathop{\rm ad} {\cal S}$ has the following form
$$
{\rm ad}\,{\cal S} \wedge \mathop{\rm ad} {\cal S }= \mathop{\rm ad} {\cal S} + \nu + \stackrel{\ast}{\nu}
\quad \qquad \mbox{for}\, S = A_n,
$$
$$
\qquad \; \mathop{\rm ad} {\cal S} \wedge \mathop{\rm ad} {\cal S} = \mathop{\rm ad} {\cal S} + \mu    \qquad
\mbox {otherwise.}
\eqno(9)
$$
Here $\mu$ and $\nu$ stand for irreps different from the adjoint.  This
decomposition is a generalization of the well known theorem which states
that $\mathop{\rm ad} {\cal S}$ always appears in $\mathop{\rm ad} {\cal S} \wedge \mathop{\rm ad} {\cal S}$
\cite{sla}.  Before presenting the explicit formula for the antisymmetric
contorsion form we will make two remarks. We note that the structure of the
Lie algebra of the reductive spaces admits two natural intertwining operators
$$
\phi(X\wedge Y) = [X,Y]_{\goth m}\quad \mbox{and}
\quad \psi(X\wedge Y) = [X,Y]_{\goth h},\quad
X,Y \in {\goth m},
$$
and we introduce the mapping $j$: $\goth h \rightarrow \goth m$.

So, now it is clear that the contorsion form $\bar\beta_{as}$ (which
accordingly to Schur's lemma must be proportional to the operator $\phi$ or
$j \circ \psi$) can be written  as
$$
\bar\beta_{as}(\tilde X\wedge \tilde Y) = \frac{f(x)}{2}\,[\tilde X,\tilde
Y], \, \quad \qquad \mbox{for}\; (\pm) \;\mbox{connection},
$$
$$
\bar\beta_{as}(\tilde X\wedge \tilde Y) = 2f(x)\circ j\Bigl([\tilde X,\tilde
Y]\Bigr), \quad \mbox{for}\; (0) \;\mbox{connection}
\eqno(10)
$$
where $f(x)$ is an arbitrary scalar function on $M^{(4)}$.

Let us define  ${\rm ad}\,K$-invariant symmetric bilinear form $B(\cdot\,,\,\cdot)$
on $\goth k$
$$
B(\tilde  X,\tilde  Y) = \langle X_1,Y_1\rangle + \langle X_2,Y_2 \rangle,
\qquad \tilde  X = (X_1,X_2),\quad \tilde Y=(Y_1,Y_2).
$$
Here $\langle \cdot,\cdot \rangle$ denotes an $ad \,S$-invariant symmetric
bilinear form on $\cal S$. Inserting the expression for $\beta_{as}$ into (7)
we see that r.h.s. vanishes identically, therefore $\beta_{s}=0$.  Thus, the
invariant connections with torsion on group manifolds form a one parametric
family given by
$$
\Lambda(\tilde  X)\tilde  Y = \frac{1+f}{2} [\tilde  X,\tilde  Y]_{\goth m},
\, \quad \mbox{for}\; (\pm) \; \mbox{connection}, \eqno(11)
$$
$$
\Lambda(\tilde  X)\tilde  Y = 2f\circ j\Bigl([\tilde  X,\tilde  Y]_{\goth
h}\Bigr), \quad \mbox{for}\; (0) \; \mbox{connection}.
\eqno(12)
$$
notice that the result for $(0)$ connection, when the group manifold is
realized as a symmetric homogeneous space, differs from that obtained in
\cite{ric} for simply connected irreducible symmetric spaces. In the latter
case the only invariant connection is the  Levi-Civita one.

Introducing the homomorphism $i: \cal S \rightarrow \goth m$,\, $i(X) =
\tilde X$ and having in mind that $i(R(X_k,X_p) X_j) = \tilde  R(\tilde
X_k,\tilde X_p)\tilde  X_j $ where $\tilde R$ is the curvature tensor on
$K/H$, we get from formula (5)
$$
R_o(X_k,X_p) X_j = F(x)\;[[X_k,X_p] X_j] , \quad X_i \in {\cal S}, \quad
F(x) = \frac{f^2(x) -1}{2},
$$
which yields the following curvature tensor components
$$ R_{ijkp} =  F(x)\, C^a_{kp}\,C^b_{aj}\,g(\tilde X_b,\tilde X_i).
\eqno(13)
$$
Here $C^a_{kp}$ are the structure constants of the Lie algebra ${\cal S}$.

Analogously, we obtain from formula (3) that
$$
T_o(X_i,X_j) =
f(x)\; [X_i, X_j]
$$
and, therefore we have for the torsion components
$$ T^k_{ij} =
f(x)\;C^k_{ij}.
\eqno(14)
$$

We will use expressions (13) and (14) in the next section.

\section{Dimensional reduction}

We are interested in a model of  multidimensional gravity in $D=4+d$
dimensions, with curvature squared terms and with torsion on the
space of extra dimension.

Let $\{\hat\theta^A\}, A=1,\dots , D$ be a basis of orthonormal 1-forms on
$E$
%where $\{\theta^{\alpha}\}$ is a basis of 1-forms on $M^{(4)}$  and
%$\{\theta^a\}$, a basis of 1-forms on $S$. In this basis the metric has the
%following form $\hat g = g_{AB} \,\hat\theta^A\otimes\hat\theta^B$.
and $\hat\omega^A_{\,B}$ a metric connection 1-form which is invariant
under the action of the symmetry group $K$.  The curvature 2-form is
constructed from the connection form $\hat\omega^A_{\,B}$ by the formula
$$
\hat\Omega^A_{\,B} = d\hat\omega^A_{\,B} + \hat\omega^A_{\,C}\wedge
\hat\omega^C_{\,B} \eqno(15)
$$
and is related to the Riemann curvature tensor
$$
\hat\Omega^A_{\,B}=\frac{1}{2}R^A_{BCD}\hat\theta^C\otimes\hat\theta^D.
\eqno(16)
$$

The action we consider is of the form
$$
S= \int_E\biggl(\hat\lambda_0 {\cal L}_0 +  \hat\lambda_1 {\cal L}_1+
\hat\lambda_2 {\cal L}_2 \biggr), \eqno(17)
$$
where $\hat\lambda_n$ are multidimensional coupling constants and
$$
{\cal L}_n= \hat\Omega^{A_1 B_1} \wedge\dots \wedge \hat\Omega^{A_n
B_n}\wedge \epsilon_{A_1 B_1 \dots A_n B_n}, \qquad 2n \leq D.
$$
$$
\epsilon_{A_1 \dots A_n}= \frac{1}{(D-n)!}\, \epsilon_{A_1 \dots A_n A_{n+1}
\dots A_D}\hat\theta^{A_{n+1}}\wedge \dots \wedge \hat\theta^{A_{D}}.
$$
In usual tensor notations we have for ${\cal L}_k$
$$
{\cal L}_0=d^{4}x\, d^{d}\,\xi\sqrt{- \hat g}\; - \;\mbox {volume},
\eqno(18)
$$
$$
{\cal L}_1 =  d^{4}x\, d^{d}\,\xi\sqrt{- \hat g}\;R\; - \;\mbox {the Einstein
Lagrangian}, \eqno(19)
$$
$$
{\cal L}_2 = d^{4}x\, d^{d}\,\xi\sqrt{- \hat g}\;(R_{ABCD}\,R^{CDAB} - 4R_{AB}\,R^{BA}
+ R^2)\; - \;\mbox {the Gauss-Bonnet term}.  \eqno(20)
$$
Here $R$ is the scalar curvature,  $R_{AB}$ is Ricci tensor, the first four
coordinates are labeled by $x$, and the remaining internal coordinates by
$\xi$.  We recall that in the case when $D=4$ the Gauss-Bonnet term is
proportional to the Euler form, therefore it does not contribute to the
equations of motion.

We assume that the multidimensional space-time has the structure
$E=M^{(4)}\times S$, with $M^{(4)}$ being the Robertson-Walker space-time and
$S$ being the internal compact group manifold. We suppose that the field
equations have a vacuum solution with a symmetry $P\times K:\; P$ is the
symmetry group of the 4-dimensional space-time and $K$ is a group of
transformation of the internal coordinates.  In this case the metric of the
vacuum solution can be taken in the form (we will use the greek subindices for
the 4-dimensional space-time and the latin ones for the internal space)
$$
\hat g_{MN}=\pmatrix{g_{\mu\nu}(x)&0 \cr 0&g_{mn}(x,\,\xi) }, \qquad
g_{mn}(x,\,\xi)=\frac{L^2(x)}{L_0^2}\,\theta_m^a(\xi)\,\theta_n^b(\xi)
\,\gamma_{ab}
\eqno(21)
$$
with $g_{\mu\nu}$ being a 4-dimensional metric on $M^{(4)}$ and $g_{mn}$
being the metric of the internal space.  $L(x)$ is the radius of the internal
space $S$ and $L_0$ is some constant (which will be discussed later).
$\gamma_{ab}$ is a flat metric at the origin $o,\;\theta^m_a(\xi)$ are the
vielbeins.

We are interested in the case when there is torsion on the internal space.
So, we choose the following ansatz for the $K$-invariant connection form:
$$
\hat \omega^A_{\,B}=\pmatrix{\stackrel{\circ}{\omega}^{\alpha}_{\,\beta}&
\stackrel{\circ}{\omega}^{\alpha}_{\,b}  \cr
\stackrel{\circ}{\omega}^{a}_{\,\beta}&\stackrel{\circ}{\omega}^a_{\,b} +
\;\bar\omega^a_{\,b} }. \qquad \eqno(22)
$$
Here, as before, the small circle denotes the Levi-Civita connection and
$\bar\omega^a_{\,b}$ stands for the contorsion form already introduced above.

Our next step is to calculate the curvature tensor components (16).  They can
be found using the metricity condition
$$
d\hat g_{AB} - \hat\omega^C_{\,B}\, g_{AC} - \hat\omega^C_{\,A}\, g_{CB}=0
\eqno(23)
$$
and the structure equation
$$
\hat T^A = d\hat\theta^A + \hat\omega^A_{\,B}\wedge\hat\theta^B,
\eqno(24)
$$
where
$$
T^{\alpha} =0,\qquad T^a = f\,C^a_{bc}\,\theta^b\wedge\theta^c.
$$

After some straightforward and rather tedious calculations we obtain the
following expressions for the connection form
$$
\stackrel{\circ}{\omega}^{\alpha}_{\,\beta}=
\Gamma^{\alpha}_{\nu\beta}\theta^{\nu}, \qquad
\stackrel{\circ}{\omega}^{\alpha}_{\,b}=-\eta_{bd}\,\Bigl(\frac{L}{L_0}\Bigr)
\partial^{\alpha}\Bigl(\frac{L(x)}{L_0}\Bigr)\,\theta^d,
$$
$$
\stackrel{\circ}{\omega}^{a}_{\,\beta}=
\partial_{\beta}\ln\Bigl(\frac{L(x)}{L_0}\Bigr)\,\theta^a,\qquad
\stackrel{\circ}{\omega}^a_{\,b} =
\delta^a_b\,d\,\ln\Bigr(\frac{L}{L_0}\Bigr),\qquad \bar\omega^a_{\,b}=
\frac{f+1}{2}C^a_{db}\theta^d.  \eqno(25)
$$
where $\Gamma^{\alpha}_{\nu\beta}$ is the Christoffel symbols for the
Robertson-Walker metric.

Using formulas (15) and (16) we obtain the following values for the curvature
tensor components $R^A_{BCD}$.
$$
R_{\beta\gamma g}^{\alpha} =  R_{\beta f g}^{\alpha} =
R_{b \gamma \delta}^{\alpha} = R_{\beta \gamma_\delta}^{a}=
R_{b \gamma \delta}^{a}=0,
$$
$$
R_{b\gamma g}^\alpha = -\frac{L(x)}{L_0}\,\nabla_\gamma
\Bigl(\partial^\alpha \frac{L(x)}{L_0}\Bigr)\,\eta_{bd}, \qquad
R_{\beta\gamma g}^a=\delta_g^a\frac{L_0}{L(x)}\,\nabla_\gamma
\Bigl(\partial_\beta \frac{L(x)}{L_0}\Bigr),
$$
$$
R_{b\gamma g}^a = \frac{\partial_\gamma f(x)}{2}\,C_{gb}^a,\qquad
R_{\beta fg}^a=f(x)\,\partial_\beta\ln \frac{L(x)}{L_0}C_{fg}^a,
$$
$$
R_{b f g}^{\alpha} = -f(x)\;\frac{L(x)}{L_0}\partial^\alpha\, \biggl(
\frac{L(x)}{L_0}\biggr)\, C^d_{gb}\,\eta_{df} ,\qquad R_{\beta f g}^{\alpha}=
f(x)\partial_{\beta}\ln \frac{L(x)}{L_0}C^a_{fg},
$$
$$
R_{b f d}^{a}=  F(x)\, C^a_{db} C^d_{fg} - \Bigl(\partial
\frac{L(x)}{L_0}\Bigr)^2 \Bigl(\delta_f^a \eta_{bg} -\delta^a_g
\eta_{bf}\Bigr),
\eqno(26)
$$
where
$$
\qquad F(x) = \frac{f^2(x) -1}{2}.
$$

For the latter use it is convenient to introduce the so-called dilaton field
$\Psi $ defined by
$$
\Psi=\ln\left\{\frac{L(x)}{L_0}\right\}.
\eqno(27)
$$

When calculating the curvature $R$ and the Gauss-Bonnet term ${\cal R}^2$ we
are faced with products of the structure constants. We can express them
through the eigenvalues of the second order Casimir operator $C_2$.  Using
that  $C_2=1$  for the adjoint representation \cite{cor}, we get
$$
R=R^{(4)}-2d\,\Box\Psi(x)-d\,(d+1)\,(\partial\Psi(x))^2- d F(x)L^{-2}(x),
\eqno(28)
$$
$$
{\cal R}^2={\cal R}^{(4)2}+8d\,(\nabla_{\nu\mu}\Psi+
\partial_\mu\Psi\,\partial_\nu\Psi)\,R^{(4)\mu\nu}-
2d\,\Bigl[2\,\Box\Psi+(d+1)(\partial\Psi)^2+ FL^{-2}\Bigr]\,R^{(4)}
$$
$$
-4d\,(d-1)\,\nabla_{\mu\nu}\Psi\,\nabla^{\mu\nu}\Psi-8d\,(d-1)\,
\nabla_{\mu\nu}\Psi\,\partial^\mu\Psi\,\partial^\nu\Psi+d\,(d-1)
(d-2)(d+1)(\partial\Psi)^4
$$
$$
+4d\,(d-1)(\Box\Psi)^2+d\,(d-3) F^2L^{-4}+4d\,(d-2) FL^{-2}
\,\Box\Psi
$$
$$
+2 Fd\,(d-1)(d-2)L^{-2}(\partial\Psi)^2+4d^2(d-1)(\partial\Psi)^2
\,\Box\Psi+2d\,f\,\partial_\mu f\,\partial^\mu\Psi L^{-2}.
\eqno(28')
$$

The invariance of the metric and the connection allows us to carry out the
dimensional reduction of the action. We insert expressions (28) and ($28'$)
into (17) and omit total derivatives. Then, after having integrated the
Lagrangian over the group manifold, we obtain the reduced action in the form
\goodbreak
$$
S=v_d\int d^4x\,e^{d\,\Psi}\sqrt{-g^{(4)}}\left\{
\hat\lambda_0+\hat\lambda_1\left[R^{(4)}+d\,(d-1)
(\partial\Psi)^2-
d FL^{-2}\right]\right.+
$$
$$
+\hat\lambda_2\Bigl[{\cal R}^{(4)2}-2d\,(d-1)(d-2)\,
\Box\Psi(\partial\Psi)^2-d\,(d-1)^2(d-2)(\partial\Psi)^4
$$
$$
-2d\,(d-2)(d-3) FL^{-2}(\partial \Psi)^2+d\,(d-3) F^2
L^{-4}-4d\,(d-3)\,\partial_\mu F\,\partial^\mu\Psi L^{-2}
$$
$$
-4d\,(d-1) G_{\mu\nu}\,\partial^\mu\Psi\,\partial^\nu\Psi-
2d R^{(4)} FL^{-2}\Bigr]\Bigr\},
\eqno(29)
$$
where $v_d$ is the volume of the internal space with the scale factor $L=L_0,\;
g^{(4)} =det\;g_{\mu\nu}$, and $G_{\mu\nu}$ is the Einstein tensor.

Notice that in the action we obtained the term proportional to $R^{(4)}$
should correspond to the Einstein gravity; therefore, to bring the action
(29) to the correct form we will introduce the metric $\eta_{\alpha\beta}(x)$
related to $g_{\alpha\beta}(x)$ by the formula
$$
g_{\alpha\beta}(x) =
\left(\frac{L(x)}{L_0}\right)^{-d}\,\eta_{\alpha\beta}\,(x).
$$

In terms of the metric
$\eta_{\alpha\beta}(x)$ the action (29) reads
$$
S=\int d^4x\sqrt{-\eta} \left\{
\bar\lambda_1R^{(4)}+\bar\lambda_0\,e^{-\Psi d}+\bar\lambda_1
\bigl(-\frac{1}{2}\,d(d+2)(\partial\Psi)^2-e^{-(d+2)\Psi}\,\bar Fd
\bigr)\right.
$$
$$
+\bar\lambda_2\,\Bigl[e^{-2\Psi}\bigl\{d(d^2-2d-12)\,\bar
F\,(\partial\Psi)^2+2d (d+6)\,\partial_\mu {\bar F}\,\partial^\mu\Psi-2d\bar
FR^{(4)}\bigr\}
$$
$$
+\,e^{d\,\Psi}\bigl\{{{\cal
R}^{(4)}}^2-d(d-1)(d+2)(\partial\Psi)^4+d(d^2-4)\,\Box\Psi\,
(\partial\Psi)^2+4dG_{\mu\nu}\,\partial^\mu\Psi\,\partial^\nu\Psi\bigr\}
$$
$$
+ \left. \left. e^{-(d+4)\Psi}d(d-3)\bar F^2\right]\right\},
\eqno(30)
$$
where $\bar\lambda_i=v^d\,\hat\lambda_i$ are the effective four-dimensional
coupling constants; $ \bar F= F\,L_0^{-2};\; R^{(4)}$, ${\cal R}^{(4)}$ and
$G_{\mu\nu}$ have been calculated with respect to the metric $\eta_{\mu\nu}$.

We consider  the spatially homogeneous and isotropic Robertson-Walker
universe.  The interval for the metric $\eta_{\alpha\beta}$ in the coordinate
basis reads
$$
ds^2_{RW} = -dt^2 + a^2(t)\biggl[d\chi^2+\sin^2\chi\bigl(d\vartheta^2
+\sin^2\vartheta\,d\phi^2\bigl)\biggl].  \eqno(31)
$$
In Robertson-Walker cosmology the function $a(t)$ is usually interpreted as
the radius of the universe. Notice that the flat Minkowski universe is the
limiting case $a \to \infty$.

For the latter use it is convenient to define the following dimensionless
variable
$$
a(t)/L_0 \equiv e^{\alpha(t)}.
\eqno(32)
$$

Straightforward calculations of the curvature $R^{(4)}$ of the
Robertson-Walker universe give the following result
$$
R^{(4)}=6(\ddot \alpha +2\dot \alpha^2)+ \frac{e^{-2\alpha}}{L_0^2}.
\eqno(33)
$$

For the scalar fields $\Psi$ and $f$ to be consistent with the homogeneous
and isotropic Robertson-Walker  ansatz, which has been chosen  for
the metric $\eta_{\mu\nu}$, they can depend on time only.
So, now the action is
$$
S=\int d^4x\sqrt{-\eta}\Bigl\{
\bar\lambda_1R^{(4)}+\bar\lambda_1\bigl( \frac{d}{2}(d+2)\dot\Psi^2\bigr)
$$
$$
+\bar\lambda_2\Bigl[\,e^{-2\Psi}\bigl\{-d\bar F\bigl((d^2-2d-12)\bar
F\dot\Psi^2+2\Psi\dot\Psi \dot\alpha-12\dot\alpha^2) +2d\dot {\bar
F}(6\dot\alpha-(d+6)\dot\Psi\bigr)\bigr\}
$$
$$
+\,e^{d\,\Psi}\bigl\{-\frac{1}{3}\,d(d+2)(d^2+d-3)\dot\Psi^4+
2d(d^2-4)\dot\Psi^3\dot\alpha+4d\dot\Psi^2(3\dot\alpha^2
+\frac{1}{2}\,e^{-2\alpha})
$$
$$
-8d\dot\Psi\dot\alpha^3- 4dR^{(3)}e^{-2\alpha}\dot\alpha\dot\Psi
\bigr\}\Bigl]-
W(\Psi,\,f,\,\alpha;\,d,\,\bar\lambda_0,\,\bar\lambda_1\,\bar\lambda_2)\Bigr\},
\eqno(34)
$$
with
$$
W=e^{-\Psi d}\{-\bar\lambda_0 + \bar\lambda_1\,d\,\bar Fe^{-2\Psi}
-\bar\lambda_2\,d(d-3)\,\bar F^2e^{-4\Psi}\} +
\frac{2d}{L_0^2}\,\bar\lambda_2\;\bar F\;e^{-2\Psi}\,e^{-2\alpha}.
\eqno(35)
$$
We notice that terms proportional to $e^{-4\alpha}$, which one could
expect to be present in the potential, in fact do not appear  due to
a special combinations of the coefficients in the Gauss-Bonnet term
(20).

Thus, after the dimensional reduction we have obtained an effective
4-dimensional theory on $M^{(4)}$, which describes Einstein gravity coupled
to the scalar fields $\Psi(t)$ and $f(t)$. We will devote the next section to
a more detailed analysis of the potential (35).

\section{Analysis of the potential}

To investigate the behavior of the effective 4-dimensional system  we need to
derive equations of motion from the action (34) and to solve them.  But the
system of equations of motion which we obtain is so complicated, that we
could not find any exact solution to it without additional assumptions. So,
we restrict ourselves to the analysis of the static solutions. We think that
\begin{figure}[h]
\centering{\leavevmode\epsfbox{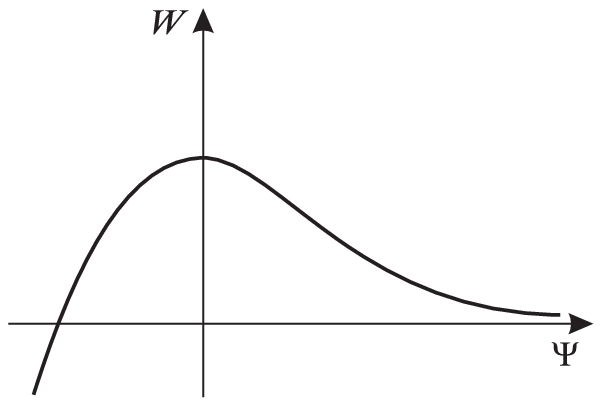}

}
\caption{}

\

\centering{\leavevmode\epsfbox{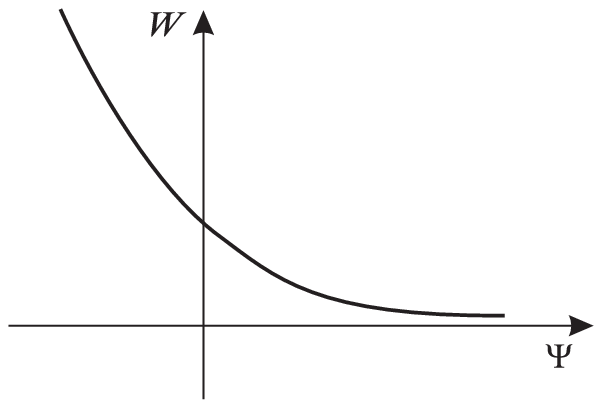}

}
\caption{}
\end{figure}
even in this particular case we can see some typical features in the
evolution of the investigated system. Static solutions of the equations of
motion correspond to the extrema of the potential (35).  The analysis of the
shape of the potential can trace qualitatively the dynamics of the internal
and the 4-dimensional scale factors, determine the values of parameters for
which the minima are separated from the decompactification area by a finite
barrier and to examine the role of curvature and torsion in these processes.
A simple analysis shows that the form of the
potential strongly depends on the values of the parameters $\bar\lambda_i$.
There are three typical shapes of the potential considered as a function of
$\Psi$.  They are shown in Fig.1 -- Fig.3.  The first two regimes are not of
physical interest, because the absence of a local minimum corresponds to the
absence of a compactifying solution with a finite size of the internal space.
If we assume that the torsion field takes an equilibrium value $f=f_{\rm
{min}}$, so in the first regime (which is the case, for example, when
$\bar\lambda_2>0, \, \bar\lambda_0<0$) the potential is unbounded from below.
Therefore, the system either enters the quantum domain (see Fig.1, negative
$\Psi$) and, therefore, the classical analysis will fail there, or $L(t)\to
\infty $ as $t\to \infty$ (for large positive $\Psi$) and we have  the
decompactification of the extra dimensions.  In the second regime (with $
\bar\lambda_0<0,\,-\frac{\bar\lambda_1^2} {4\,\bar\lambda_0 \bar\lambda_2}\,<
\frac{(d+4)(d-3)}{(d+2)^2}$) we also see that the decompactification of the
extra dimensions takes place (see Fig.2, large positive~$\Psi $).

So, in the next two subsection we will study only those values of the
parameters for which the potential has classically stable local minima,
separated from the decompactification region by a finite barrier (Fig.3).

\begin{figure}
\centering{\leavevmode\epsfbox{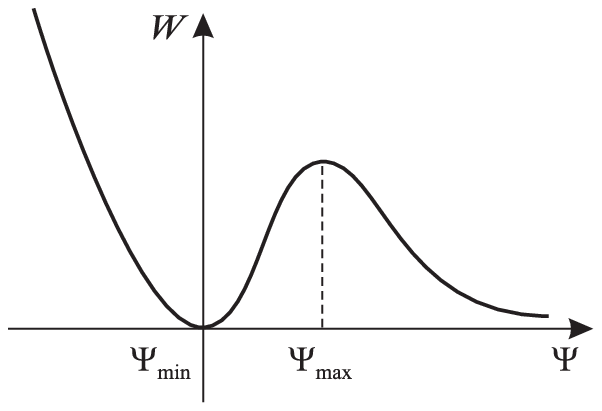}

}
\caption{}
\end{figure}

For the purpose of comparison we will use some results from our previous
paper \cite{kub}, where the 4-dimensional space-time was chosen to be the
flat Minkowski space-time (that is the case $\alpha \to \infty $). The
potential of our present model, which corresponds to the Robertson-Walker
metric, differs from the Minkowski case by the presence of the last term in
the expression (35).  This term makes the analysis much more difficult, so we
are forced to apply the method of perturbations with respect to a small
parameter  $\varepsilon$.  Using the typical estimations for the coupling
constants $\bar\lambda_i$ \cite {gut}, \cite{kol}, we  found that it is
convenient to choose $\varepsilon$ to be  $\varepsilon\equiv e^{-2\alpha}$;
where $\varepsilon$ satisfies the condition $\varepsilon\ll\bar\lambda_1\,
L_0^2$ and we also assume that $\varepsilon\ll \bar\lambda_0\, L_0^4$, and
$\varepsilon\ll \bar\lambda_2$.  In our calculations of the minima of the
potential (35) it seems to be natural to choose the corresponding quantities
of the flat Minkowski case as zero approximation.

We will study the potential as a function of two variables $W=W(\Psi,\; f)$
and neglect the fact that $W$ also depends on $\alpha$. We may assume that
$\alpha$ changes little in comparison with $\Psi$ and $ f$, so that this
assumption allows us to consider $\alpha$ as a `constant background'.

Before investigating the minima  of the potential we discuss the constant
$L_0$ which still remains a free parameter in the model. Let us consider the
flat Minkowski metric (that is the limiting case $\varepsilon =0$). We can
fix the parameter $L_0$ by the condition that the potential $W=W(\Psi)$ takes
its minimum at $\Psi=\Psi_{\rm min}=0$. This allows us to interpret $L_0$ as
the size of the internal space at the vacuum state, corresponding to the
compactification of the extra dimensions.  The explicit formula for $L_0$
will be presented in the next subsections.

We will investigate those  values of the parameters $\bar\lambda_i$ for which
there were minima of the potential in the Minkowski case, so that we will be
able to compare the results.  In what follows we will also use the notation
$\Delta\equiv\frac{d\,\bar\lambda_1^2} {4(d-3)\,\bar\lambda_0
\bar\lambda_2}$.

\subsection{Case 1: $\quad \bar\lambda_0<0,\quad \bar\lambda_2<0,\quad d\geq
4,\quad \Delta=1$}

Under these values of the parameters in the case of the
flat Minkowski metric the minima of the corresponding potential are
degenerated and located on the curve with nonzero torsion $\,|f|<1$
$$
\Psi_{\rm min}(f)=\frac{1}{2}\ln\left\{-(d-3)
\frac{\bar\lambda_2}
{\bar\lambda_1\,L_0^2}
\,\frac{1-f^2}{2}\right\}.
\eqno(36)
$$
This curve consists from two gutters, which join each other at the point
with zero torsion.

Imposing the condition $\Psi_{\rm min}=0$, as it was explained earlier, we
obtain the following value of the parameter $L_0$
$$
L_0=\sqrt{\frac{-\bar\lambda_2\,(d-3)}{2\bar\lambda_1}}.
\eqno(37)
$$
Looking at formula (34) we see that the value of the potential in the minimum
generates an effective cosmological constant $\Lambda^{(4)}= W(\Psi_{\rm
min},f_{\rm min})$ in four dimensions.  It can be checked that under this
values of the parameters the potential vanishes identically on the curve
(36), therefore the four-dimensional cosmological constant
$\Lambda^{(4)}_{\rm Min}$ is equal to zero, as it was expected.

When we pass to the Robertson-Walker metric we find that if the curvature is
present  the curve of the minima (36) with nonzero torsion reduces to the
only point and we are left with the  only minimum corresponding to zero
torsion
$$
\left(\Psi_{\rm min}=-\frac{1}{2}\ln\left\{1+2\frac{\bar\lambda_2}
{\bar\lambda_1\,L_0^2}\,\varepsilon\right\},\ f_{\rm min}=0\right),
\eqno(38)
$$
where we had to apply the second order of the theory of perturbation to see
that the degeneracy is removed and the previous gutters ascend and no longer
correspond to minima.  Now we obtain a nonzero value for the effective four
dimensional cosmological constant $\Lambda^{(4)}_{\rm RW}$
$$
\Lambda^{(4)}_{\rm RW}\sim
\frac{d}{d-3}\,\frac{\bar\lambda_1}{L_0^2}\,\varepsilon.
$$
We see that $\Lambda^{(4)}_{\rm RW}$ is given by the well defined quantities:
$\varepsilon$ (related to the curvature), $\bar\lambda_1$ (which is
the Newton constant) and $d$ (the dimension of the internal space).
Thus, there are no free parameters which we could turne
with the aim to set $\Lambda^{(4)}_{\rm RW}$ equal to zero.

\subsection{Case 2: $\quad \bar\lambda_0<0,\quad \bar\lambda_2<0,
\quad d\geq 4,\quad\frac{d(d+4)}{(d+2)^2} < \Delta <1$}

In the  flat Minkowski case we found the there was a minimum with zero
torsion for these values of the parameters.  For $\varepsilon\neq 0$ our
calculations within the perturbation theory shows that, as in the previous
case, the potential (35) has a minimum only when torsion is zero.  This
minimum is located at the point
$$
\left(\Psi_{\rm min}=-\frac{1}{2}\ln\left[1+
\frac{4}{H\,(d+2)}\,\frac{\bar\lambda_2}{\bar\lambda_1\,L_0^2}\,
\varepsilon\right],\ f_{\rm min}=0\right)
\eqno(39)
$$
where
$H\equiv\sqrt{1-\frac{D(d)}{\Delta}},\;D(d)\equiv\frac{d(d+4)}{(d+2)^2}$.

In this case we obtain the following expression for $L_0$
$$
L_0=\frac{\bar\lambda_1}{\bar\lambda_0}\,\frac{(d+2)}{8(H-1)},
\eqno(40)
$$
which turns to the form (37) for $\Delta=1$.

As in the previous subsection, we also obtain a nonvanishing four dimensional
cosmological constant, shifted by the presence of the curvature; but this
expression is too lengthy, so we will not present it here.

\section{Conclusions}

The analysis we made shows that the curvature of the Robertson-Walker
space-time plays an important role and essentially affects the model. In
contrast to the flat Minkowski metric for non-vanishing curvature there are
no longer minima with non-zero torsion. Let us suppose that the initial
conditions for scalar fields $\Psi,\; f$ and for their derivatives
$\dot\Psi,\; \dot f$ were such that the universe has reached the minimum and
remained there without overcoming the barrier of the potential (see Fig.3).
Such a situation corresponds  to the compactification of the internal space
with the characteristic size given by (37) or (40). So, in this case we see
that if the curvature of the 4-dimensional space is nonzero,  there is no
solution of spontaneous compactification with nonzero torsion on the internal
space. This is in contrast with the case of the Minkowski space-time, where
minima of the potential with nonzero torsion do exist.

We found that the value of the potential for the spontaneous compactification
minimum is non zero, which immediately implies that the vacuum state is
stable only classically. So, in this case  there is the possibility for
decompactification of the internal space via quantum tunneling through the
potential barrier. However, under some general assumptions about the values
of the parameters of the model, the time of such a tunneling exceeds the
life-time of the universe.

\section*{Acknowledgments}
The authors are very thankful to Yu.A.\,Kubyshin, I.P.\,Volobuev and
J.I.\,P\'erez Cadenas for illuminating discussions.

%\begin{figure}
%\centering
%\unitlength=1mm
%\begin{picture}(80,50)
%\put(10,50){
%\special{em:graph article1.pcx}
%}
%\end{picture}
%\caption{}
%%{Fig.1}
%\end{figure}

%\begin{figure}
%\centering
%\unitlength=1mm
%\begin{picture}(80,50)
%\put(10,50){
%\special{em:graph article2.pcx}
%}
%\end{picture}
%\caption{}
%\end{figure}
%\begin{figure}
%\centering
%\unitlength=1mm
%\begin{picture}(80,50)
%\put(10,50){
%\special{em:graph article3.pcx}
%}
%\end{picture}
%\caption{}
%% }
%\end{figure}

%            REFERENCES

\end{document}